\documentclass[%
 reprint,
superscriptaddress,
 amsmath,amssymb,
 aps,prl,
]{revtex4-1}

\usepackage{hyperref}
\hypersetup{
colorlinks=true,
linkcolor=black,
filecolor=blue,
citecolor=blue,  
urlcolor=black,
}

\usepackage{color} 
\usepackage{graphicx}
\usepackage{dcolumn}
\usepackage{bm}

\usepackage{verbatim}
\usepackage{appendix}
\usepackage{amsthm}
\theoremstyle{plain}
\newtheorem{thm}{Theorem}

\theoremstyle{definition}

\begin{document}

\title{No energy transport without discord}
\author{Seth Lloyd}
\affiliation{%
Department of Mechanical Engineering, Massachusetts Institute of Technology, Cambridge, Massachusetts 02139, USA}
\affiliation{%
Department of Physics, Massachusetts Institute of Technology, Cambridge, Massachusetts 02139, USA}
\author{Zi-Wen Liu}
\affiliation{%
Center for Theoretical Physics, Massachusetts Institute of Technology, Cambridge, Massachusetts 02139, USA}
\affiliation{%
Department of Physics, Massachusetts Institute of Technology, Cambridge, Massachusetts 02139, USA}
\author{Stefano Pirandola}
\affiliation{%
Computer Science and York Centre for Quantum Technologies,
University of York, York YO10 5GH, UK}
\author{Vazrik Chiloyan}
\affiliation{%
Department of Mechanical Engineering, Massachusetts Institute of Technology, Cambridge, Massachusetts 02139, USA}
\author{Yongjie Hu}
\affiliation{Department of Mechanical and Aerospace Engineering, University of California, Los Angeles, California 90095, USA}
\author{Samuel Huberman}
\affiliation{%
Department of Mechanical Engineering, Massachusetts Institute of Technology, Cambridge, Massachusetts 02139, USA}

\author{Gang Chen}
\affiliation{%
Department of Mechanical Engineering, Massachusetts Institute of Technology, Cambridge, Massachusetts 02139, USA}

\date{\today}


\begin{abstract}
We show that without quantum correlations, energy cannot flow. The result follows from a simple theorem that shows that systems whose dynamics do not generate quantum discord are effectively non-interacting.
We show that the rate of heat transfer between two quantum
systems at different temperatures is directly proportional
to the instantaneous rate of increase of diagonal/energetic discord between the systems.  Consequently, nanoscale heat transfer experiments
can be used to measure discord directly.
We report the results of a measurement
of the increase in discord due to nanoscale heat flow across an
aluminum-sapphire interface and find it to be $4.28
\times 10^{24}$ ${\rm bits}\,{\rm m^{-2}}\,{\rm K^{-1}}\,{\rm s^{-1}}$.
\end{abstract}

\maketitle

Quantum systems can be correlated in ways that classical systems can not. 
A wide variety of non-classical forms of correlation exist \cite{nc,zurek,zurekmd,hv,ohhh,phh,wpm,mpsvw,datta,dg,luo,hjpw,bennett,lcs,ws}: amongst the
best known are entanglement \cite{nc} and discord \cite{zurek,zurekmd,hv,ohhh,datta,dg}.
Quantum correlations can be used to enhance measurement accuracy
\cite{glm}, energy transport \cite{reben,ph,goldi}, and to establish private communications \cite{Pirandola2014,Pirandola2017}. 
Discord measures the difference between quantum mutual information
and the classical mutual information between local measurement results. It aims to quantify the amount of all nonclassical correlations. (See \cite{modi} for a comprehensive review on discord.)
More formally, the ordinary discord $D(A\rightarrow B)$ measures the minimum reduction
in correlation between two quantum systems, $A$ and $B$, induced
by measurements made on $A$ 
\cite{zurek,zurekmd,hv,ohhh,datta,dg}.  Similarly, $D(B\rightarrow A)$
measures the minimum reduction in correlation induced by measurements
made on $B$.  Because of the minimization, discord can be difficult
to calculate \cite{npc}.   A natural and more easily calculated version of discord
is diagonal discord -- the
reduction in correlation when $A$ and $B$ are measured in the bases
that diagonalize the reduced density matrices for $A$ and $B$ (similar measures previously discussed in e.g.\ \cite{mid,lcs}).   
In the case of thermal states, diagonal
discord is the `energetic' discord defined by measurement
in the energy eigenbasis.  As a consequence, diagonal
discord has a direct interpretation in terms of flows of energy
and entropy, and can be determined directly from measurement.

Evolutions that do not create discord are regarded as classical. For instance, it has been shown that a large class of quantum computation that contains no discord within the register at all steps admits efficient classical simulation \cite{eastin,cable}.
In this paper, we show that quantum correlations -- in the form
of discord -- are mandatory for {\it any}
energy transport taking place between parts of a closed system.  
That is, energy transport cannot occur between subsystems that remain classically correlated (discord-free) during the evolution.
Moreover, we show that the initial rate of heat transfer between two
systems prepared at different temperatures is proportional to
the rate of increase in diagonal/energetic discord between the systems. When the systems are in a Gaussian state, we show that we can compare diagonal discord to the actual discord, and that they converge in the high-temperature limit. This result allows us to measure discord directly by quantifying the instantaneous
rate of heat flow between two systems, without having to perform detailed state tomography.  Effectively, the heat flow across the surface acts as a witness of the presence of discord, and allows us to measure it.  We exhibit an experiment in which we measure the increase of energetic discord induced by nanoscale heat flow across an aluminum-sapphire
interface when the aluminum is excited by a femtosecond laser pulse.  The rate of increase of discord is a function only of the temperature difference and
heat flux across the interface, and is measured to be
$4.28
\times 10^{24}$ ${\rm bits}\,{\rm m^{-2}}\,{\rm K^{-1}}\,{\rm s^{-1}}$.
To our knowledge, this is the first experiment to demonstrate and certify theoretically the creation of macroscopic amounts of discord.  Note that
the experiment is not a test of a theoretical prediction, but instead a direct measurement of discord based on the theory.  Our theoretical results consist not of predictions of the results of experiments, but rather a set of theorems that allow us to relate quantity of discord to experimentally measurable quantities.  We prove that instantaneous heat flow is a `discord witness,' and that the amount of discord created is a function of the measured amount of heat flow, together with the temperatures of the substrates between which the flow takes place.

\section{Ordinary and diagonal discords}
To define discord,
consider a joint quantum system $AB$ described by a density matrix
$\rho_{AB}$.  The systems $A$ and $B$ considered individually are
described by reduced density matrices $\rho_A = {\rm tr}_B \rho_{AB}$,
$\rho_B = {\rm tr}_A \rho_{AB}$ respectively, where ${\rm tr}_{A,B}$
represents the partial trace over $A$,$B$.   The quantum mutual information
$I(A:B) = I(A) - I(A|B)$ is defined to be the difference between
the information of $A$: $I(A) =  -{\rm tr} \rho_A \ln \rho_A$ and
the quantum conditional information
$I(A|B) =
- {\rm tr} \rho_{AB} \ln \rho_{AB}
+ {\rm tr} \rho_B \ln \rho_B.$  The quantum mutual information
measures the total amount
of correlation between $A$ and $B$.  Discord compares the
quantum mutual information with the mutual information obtained
when a measurement is made locally.  Consider a measurement
made on $B$ alone.  The measurement has outcomes $b$ with
probabilities $p(b)$, and the state of $A$ given the outcome
$B$ is $\rho(A|b)$.  The mutual information between $A$ and
the results of this measurement on $B$ is equal to
$I(A:\tilde B) = I(A) - \sum_b p(b) I(A|b)$,
where $I(A|b) = -{\rm tr} \rho(A|b) \ln \rho(A|b)$.
The quantum discord is defined to be the minimum difference
between the quantum mutual information and the mutual
information given a measurement on $B$:
\begin{equation}\label{qd}
D(B\rightarrow A) = I(A:B) - \max I(A:\tilde B),
\end{equation}
where the maximum is taken over all possible measurements
on $B$.
The symmetric discord (or known as WPM discord) is the minimum difference between the quantum
mutual information and the maximum classical mutual information of
the results of measurements made on both $A$ and $B$ \cite{wpm}.

Define the diagonal discord
$D_{diag}(B\rightarrow A)$ to be the difference between
the quantum mutual information and the mutual information
given a measurement of $B$ consisting of rank one projectors
onto the states of the Schmidt basis: that is, we measure $B$ in
the basis with respect to which $\rho_B$ is diagonal. 
(If $\rho_B$ has degenerate eigenvalues we use the basis
for the degenerate subspace that minimizes the entropy increase.)
Clearly, $D_{diag}(B\rightarrow A) \geq D(B\rightarrow A)$.

Like ordinary discord, diagonal discord quantifies the reduction in mutual information
induced by local measurement, which can be regarded as the nonclassical part of correlations. Ordinary discord is defined by the optimal local measurement, but diagonal discord is much easier to calculate because we let the reduced state define the
measurement to be made. The choice of Schmidt basis is natural. It is indeed true that Schmidt basis measurement does not always capture all locally accessible (classical) correlations due to the absence of optimization. A critical issue is that the local eigenbasis is not uniquely defined when the reduced states possess degeneracy, which can lead to other subtleties, e.g., discontinuity \cite{wpm,criteria}. (It can be shown that diagonal discord and the local eigenbasis are continuous when the local spectrum is nondegenerate \cite{2017arXiv170809076L}.)
However, it can be easily seen that diagonal and ordinary discord vanish for the same set of states. Thus operations that cannot create ordinary discord also cannot create diagonal discord.
In  \cite{2017arXiv170809076L,rd}, we prove several desirable features of diagonal discord: for example, we find that diagonal discord is likely a monotone nonincreasing under all discord non-generating local channels.  Note that the diagonal discord defined on both sides is also known as measurement induced disturbance \cite{mid}.

In the case of Gaussian states, where discord can be calculated directly \cite{OptimalityDiscord}, we can compare diagonal discord to the ordinary definition.    In the appendix, we
show that the diagonal discord and ordinary discord for Gaussian states coincide in the high temperature limit.   The Gaussian, high-temperature limit is the case that is relevant for our experimental demonstration of discord creation.

In addition, diagonal discord has natural correspondences in various thermodynamical scenarios. For example, consider a scenario where two parties share a joint state, but they only have knowledge of their own local states. They want to extract work from the joint state. Then diagonal discord determines the maximum extra amount of total work that they can can extract when they have access to quantum communication channels instead of classical ones \cite{dddemon}.
This paper shows that diagonal discord also has a natural physical interpretation in terms of energy flow and entropy increase, and can
be measured directly by experiment.   

\section{No energy transport without discord}

Our first result is that the presence of discord -- ordinary
or diagonal -- is a
prerequisite for {\it any} non-trivial interaction
between subsystems of a closed system.  More explicitly, we prove the following theorem:
\begin{thm}
Consider two quantum
systems $A$ and $B$ evolving continuously from an initial
state $\rho_{AB}(0)$ under a unitary time evolution $U_{AB}(t)$.
If the discord from
$A$ to $B$ and from $B$ to $A$ is zero for all times $t$,
then the time evolution for $A$ and $B$ can always be written as
\begin{equation}\label{non}
\rho_{AB}(t) = U_A(t) \otimes U_B(t)~ \rho_{AB}(0)~
U^\dagger_A(t) \otimes U^\dagger_B(t).
\end{equation}
In other words, a system
with zero double-way discord is effectively non-interacting.
\end{thm}

\begin{proof}

Zero discord from $A$ to $B$,
$D(A \rightarrow B) = 0$, implies that
\begin{equation}\label{zeroa}
\rho_{AB}(t) = \sum_{j} p_j(t)
|j(t)\rangle_A\langle j(t)|
\otimes \rho^j_B(t),
\end{equation}
where $|j(t)\rangle_A$ is the Schmidt
basis for $A$ at time $t$, i.e., the basis with respect to which
$\rho_A(t)$ is diagonal, and $\rho_B^j(t)$ is a density matrix
for $B$ \cite{dg}.  Similarly, zero discord
from $B$ to $A$ implies that
\begin{equation}\label{zerob}
\rho_{AB}(t) = \sum_{k} p_k(t)
\rho^k_{A}(t)\otimes
|k(t)\rangle_B\langle k(t)|,
\end{equation}
where $|k(t)\rangle_B$ is the Schmidt basis for $B$.

Zero discord from
$A$ to $B$ and from $B$ to $A$ imply that measurement of both $A$ and $B$ in
their local eigenbases (Schmidt bases) leaves $\rho_{AB}$ unchanged.
Equations (\ref{zeroa},\ref{zerob}) then immediately imply that 
$\rho_B^j(t)$
is diagonal in the Schmidt basis for $B$, for all $j$, and
$\rho_A^k(t)$ is diagonal in the Schmidt basis for $A$.
Equivalently, 
under a joint unitary evolution $U_{AB}(t)$ continuously connected
to the identity at $t=0$, 
the density matrix at all times $t$ can be written as
\begin{eqnarray}\label{dyn}
\rho_{AB}(t) &=& \sum_{jk} p_{jk}(t) |j(t)\rangle_A\langle j(t)|
\otimes |k(t)\rangle_B\langle k(t)|\nonumber\\
&=& U_A(t)\otimes U_B(t) ~ \rho_{AB}(0) ~
U_A^\dagger(t)\otimes U_B^\dagger(t),
\end{eqnarray}
where $U_A(t)$, $U_B(t)$ transform the Schmidt bases of $A$,$B$
at time $0$ to time $t$: $|j(t)\rangle_A =  U_A(t) |j(0)\rangle_A$,
$|k(t)\rangle_B = U_B(t) |k(0)\rangle_B$.
Here we have used the fact that since $U_{AB}(t)$ is continuously
connected to the identity
and preserves the eigenvalues $p_{jk}(t)$ of $\rho_{AB}$,
$p_{jk}(t) = p_{jk}(0)$ remain unchanged during the evolution. Note that this does not necessarily hold when the joint evolution is not unitary. That is, our result only applies to closed system dynamics.
This proves the theorem:
zero symmetric discord implies that $A$ and $B$ are effectively
uncoupled.
\end{proof}

Equation (\ref{dyn})  implies that, to preserve zero symmetric discord,
the effective Hamiltonian
for $A$ and $B$ must be of the form $H = \tilde H_A \otimes I_B +
I_A \otimes \tilde H_B$.  Indeed, for short times $\Delta t$, write
$U_A(\Delta  t) = e^{-i\tilde H_A\Delta t}$, and
$U_B(\Delta  t) = e^{-i\tilde H_B\Delta t}$.   Expanding to
first order in $\Delta t$, we have
\begin{eqnarray}\label{ham}
&&(H_A\otimes I_B +  I_A \otimes H_B + H_{AB}) |i\rangle_A|j\rangle_B\nonumber\\
&=& (\tilde H_A \otimes I_B + \tilde I_A\otimes H_B) |i\rangle_A|j\rangle_B,
\end{eqnarray}
for all $|i\rangle_A|j\rangle_B$.  In other words,
$H_{AB} = (\tilde H_A - H_A)\otimes I_B + I_A\otimes(\tilde H_B - H_B)$, and
the two systems are effectively non-interacting.

The proof is by and large straightforward but possesses
two subtleties.  First, the theorem requires that the time
evolution evolves continuously from the identity transformation,
i.e., $U(0) = I$.  Discontinuous transformations such as the SWAP
operator that interchanges the states of $A$ and $B$ need not
generate discord, but if such transformations are the end result
of a continuous unitary time evolution, then discord must be
generated somewhere along the way.  The second subtlety
arises in the case when the initial density matrix
is degenerate, so that its eigenvalues are the same on some subspace.
In this case, the actual physical unitary
evolution can induce couplings within the degenerate subspace:
but because the subspace is degenerate, the time evolution is always
equivalent to non-interacting dynamics as in the theorem.
This implies our second result, that energy transport cannot take place in
the absence of discord.   In particular, consider two systems
$A$ and $B$ with joint Hamiltonian $H_{AB}$.
The proof of the theorem
implies that on subspaces where $\rho_{AB}$ is non-degenerate,
$H_{AB} = H_A + H_B$: there is no interaction and no energy transfer.
Conversely, within degenerate subspaces, we can have interaction,
but the degeneracy of these subspaces precludes energy transport.
There is no energy transport without discord.

\section{Heat flow is proportional to diagonal discord}

Since energy transport must generate quantum correlations, we
can ask how much discord is generated when energy is transported.
In fact, as we now show, 
when $A$ and $B$ are initially prepared in thermal
states at different temperatures, the rate of energy flow from $A$
to $B$ is directly proportional to the rate of generation of diagonal discord.

Suppose that the initial states of $A$ and $B$ are
thermal states at temperatures $T_A$, $T_B$, $T_A > T_B$.
The initial density matrix for $AB$ is
$\rho_{AB} = Z_A^{-1} e^{-H_A/kT_A} \otimes Z_B^{-1} e^{-H_B/kT_B}$, where $k$ is Boltzmann's constant.
The two systems are uncorrelated; the Schmidt bases for $A$ and $B$ are
their energy eigenbases; and the discord and diagonal
discord are zero.  Over a brief period of time $\Delta t$, $\rho_{AB}$
evolves to
$ \rho_{AB} + \Delta \rho_{AB}
= U(\Delta t) \rho_{AB} U^\dagger(\Delta t)$.
For example, we can take $U(\Delta t) =
e^{-iH\Delta t}$, where $H = H_A + H_B + H_{AB}$,
and expand the time evolution to second order in $\Delta t$
(to first order in $\Delta t$ there is no energy transfer).
Conservation of energy and weak coupling imply that
$A$ and $B$ are still diagonal in their energy eigenbases.
The increase in the diagonal discord over
time $\Delta t$ is the difference between
the quantum mutual information between $A$ and $B$
at time $\Delta t$ and the mutual information when $B$'s
energy is measured:
\begin{equation}\label{dd}
\Delta D_{diag}(B\rightarrow A) =
I_{\Delta t}(A:B) - I_{\Delta t}(A:\tilde B).
\end{equation}

To evaluate the increase in diagonal discord, use the
following facts about entropy and mutual information.
The mutual information $I(A:B)$ is equal to
$I(A) + I(B) - I(AB)$, where $I(A) = -{\rm tr } \rho_A
\log \rho_A$, $I(B) = -{\rm tr} \rho_B \log \rho_B$,
and $I(AB) = - {\rm tr} \rho_{AB} \log \rho_{AB}$.
Here, $\rho_{AB}$ is the density matrix for the two
systems taken together, and $\rho_A = {\rm tr}_B \rho_{AB}$,
$\rho_B = {\rm tr}_A \rho_{AB}$ are the reduced density
matrices for the two systems taken separately.
Define $\rho_{A\tilde B}$ and $\rho_{\tilde B} = {\rm tr}_A \rho_{A\tilde B}$
to be the density matrices
for $AB$ and for $B$ after a projective measurement of
energy has been made on $B$.  The information $I(A\tilde B)$
and mutual information $I(A:\tilde B)$ are defined
in terms of these post-measurement density matrices.

The change in discord over time $\Delta t$ is given by
\begin{eqnarray}\label{ddt}
\Delta D_{diag}(B\rightarrow A) &=&
I_{\Delta t}(A:B) - I_{\Delta t}(A:\tilde B)\nonumber\\
&=& \Delta I(A) + \Delta I(B) - \Delta I(AB)\nonumber\\
&&- \Delta I(A) - \Delta I(\tilde B) + \Delta I(A\tilde B).
\end{eqnarray}
This expression for the change in energetic discord
can be simplified.  Because the time evolution is unitary,
we have $\Delta I(AB)  = 0$.
The infinitesimal change in entropy $I = -{\rm tr}
\rho \log \rho$ when $\rho \rightarrow \rho + \Delta \rho$
is $\Delta I = - {\rm tr} \Delta \rho \log \rho$.
Using this relationship
and noting that $\rho_B$ is already diagonal in the energy eigenbasis,
we have $\Delta I(B) = \Delta I(\tilde B)$.
Substituting these relations into Eq. (\ref{ddt})
yields
\begin{equation}\label{ddt2}
\Delta D_{diag}(B\rightarrow A) = \Delta I(A\tilde B) =
-  {\rm tr} \Delta  \rho_{A\tilde B} \log \rho_{A\tilde B}.
\end{equation}
Note that the thermal form of the initial state implies that
$\rho_{AB} = \rho_{A\tilde B}$.  Substituting the thermal state into
Eq. (\ref{ddt2}) yields
\begin{eqnarray}\label{ddt3}
\Delta D_{diag}(B\rightarrow A) &=&
({\rm tr} H_A \Delta \rho_A) /k T_A + ({\rm tr} H_B \Delta \rho_B) /k T_B \nonumber\\
&=& \Delta E_A/ kT_A + \Delta E_B/ kT_B,
\end{eqnarray}
where $\Delta E_A$ is the change in energy of $A$ and
$\Delta E_B$ is the change in energy of $B$.
In the weak coupling limit, we have $\Delta E_A = -\Delta E_B$,
and so the change in diagonal discord is
\begin{equation}\label{ddt4}
\Delta D_{diag}(B\rightarrow A) = (1/kT_B - 1/kT_A) \Delta E_B,
\end{equation}
and the energy flow from $A$ to $B$ is
\begin{equation}\label{flow}
\Delta E_B = k \frac{T_A T_B}{T_A - T_B} \Delta D_{diag}(B\rightarrow A).
\end{equation}
The flow of energy between two systems at different
temperatures is proportional to the instantaneous
increase of diagonal discord.
Writing
$\beta_A = 1/kT_A = \Delta S_A/k\Delta E_A$, $\beta_B = 1/kT_B
= \Delta S_B/\Delta E_B$, we see that
$\Delta D_{diag}(B\rightarrow A) = \Delta E_B (\beta_B - \beta_A)$
is just the increase in entropy induced by the heat transfer.
(When system $B$ is at zero temperature the rate of increase
of entropy initially diverges.)
Indeed, this interpretation of the increase in diagonal discord
follows directly from its definition: in the weak coupling limit
the density matrices for $A$ and $B$ remain diagonal in the energy
eigenbasis, and diagonal discord is the
increase in the entropy of $AB$ when a measurement of energy is
made on $A$ or $B$.   

For the Gaussian case, we are able to compute the ordinary discord.  Consider two modes $A$ and $B$ in thermal states at temperatures $T_A>T_B$. We interact them through a beamsplitter with transmissivity $\eta$. We find that, in the weak coupling limit $\eta \simeq 1$ (similar as the infinitesimal evolution time),
\begin{equation}
\Delta D(B\rightarrow A) \simeq \Delta\bar{n}_{B}\left(\frac{1}{kT_B} - \frac{1+\bar{n}_{A}}{1+\bar
{n}_{B}}\frac{1}{kT_A} \right),
\end{equation}
where $\bar{n}:=1/({e^{1/{kT}}-1})$ is the mean photon number.  For $T_A - T_B \ll T_B$,  $\bar{n}_{A}\simeq \bar{n}_{B}$. So $\Delta D(B\rightarrow A) \simeq \Delta D_{diag}(B\rightarrow A)$ in the limit of small temperature difference. (Details of the above calculations are in Appendix \ref{gau}.)

\begin{figure}
\centering
\includegraphics [width=0.95\columnwidth]{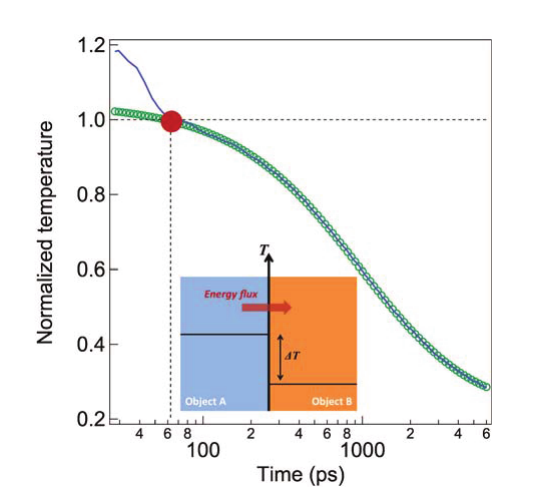}
\caption{Energy discord in thermal transport. Heat flows from aluminum 
film (object A) to sapphire substrate (object B). Experimental measurement 
within ultra-fast time domain (blue line) has been fit with Fourier 
Law (green circles). The crossing point between experiment and fitting 
is indicated with red spot, and its energy status is illustrated in the 
inset.   From Eq. (\ref{ddt4}) we can relate the rate of creation of discord
to the instantaneous energy flow between film and substrate
as a function of their difference in temperature.   Fitting to
the data, we find this rate to be at flow across an
aluminum-sapphire interface and find it to be $4.28
\times 10^{24}$ ${\rm bits}\,{\rm m^{-2}}\,{\rm K^{-1}}\,{\rm s^{-1}}$.  
}\label{Fig.1}
\end{figure}

\section{Experimental quantification of increase in discord}
Equation (\ref{flow}) shows that we do not have to perform detailed state tomography to measure discord.   Instead, we can directly determine the rate
of increase in diagonal discord by measuring ultrafast
transient heat flow in nanoscale
transport experiments \cite{vp,trev}.  Equation (\ref{ddt4}) shows that the presence of heat flow serves as a witness of the creation of discord, and allows its measurement.  Because of the Gaussian nature of the laser excitation and heat flow, we can directly compare diagonal discord to ordinary discord (see above).   As the two types of discord converge in the high-temperature limit in which the experiment took place, the experiment
also measures ordinary discord. We emphasize that the experiment is not designed to verify the predictions of the theory presented here. Nor is the experiment intended to `prove' that heat flow can't be described classically: indeed, we deduce the rate of heat flow using the Fourier law.  Instead, the theory shows that the amount of discord created between two quantum systems is directly proportional to heat flow.  The experiment measures heat flow and so measures discord.  

We measured the rate of creation of diagonal discord induced by
nanoscale heat transfer across an aluminum-sapphire
interface. 
The measurement directly determines the rate of increase
in diagonal discord
$\Delta E_B \big( 1/kT_B - 1/kT_A) =  \Delta D_{diag}(B\rightarrow A)$
per square meter per second.  
A femtosecond laser
pulse deposits energy in a thin film of aluminum on a sapphire
substrate.  In the
vicinity of the laser spot, the electrons in the
aluminum rapidly thermalize over
a time scale $t_{Th} \approx 100 ~ {\rm fs}$ that is much shorter than the time
scale $t_{tr} \approx 50-100 ~ {\rm ps}$  of initial energy transfer to the
sapphire (see Fig.~1).   The temperature of the aluminum at
the center of the spot is monitored by reflectivity as shown in
Fig.~1.  Combining the change
in temperature with the known specific heat of aluminum
allows a determination of the rate of heat transfer from the
aluminum to the sapphire substrate.

Because the film is thin ($\sim$70 nm), the primary mode of heat loss from the
spot is from propagation from the aluminum to the sapphire.  The
change in energy per unit area per unit time of the hot spot
on the aluminum is given by $\Delta E/\Delta x^2 \Delta t
= \theta C \Delta T_A/\Delta t$, where $C$ is the specific heat of aluminum
per unit volume,
$\theta$ is the film thickness, and $\Delta T_A$ is the change in the
temperature of the aluminum over time $\Delta t$.
The change in diagonal discord per unit area per unit time is then
\begin{eqnarray}
\frac{\Delta D}{\Delta x^2 \Delta t} &=& \theta C\frac{\Delta T_A}{\Delta t}
\left( \frac{1}{k T_B} - \frac{1}{k T_A} \right) \nonumber\\
&=&
\theta C \left( {\frac{T_A}{T_B} - 1} \right) \frac{1}{kT_A} \frac{\Delta T_A}{\Delta t},
\end{eqnarray}
where $T_B$ is the temperature of the sapphire.
Using the data as shown in figure 1, we obtain a value for
$\Delta D/\Delta x^2 \Delta t$ of $4.28
\times 10^{24}$ ${\rm bits}\,{\rm m^{-2}}\,{\rm K^{-1}}\,{\rm s^{-1}}$.


\section{Discussion}

This letter addressed the role played by quantum correlations in
interaction and energy transport.    We showed that quantum
correlations in the form of discord are mandatory for
energy transport between quantum systems.    We defined
a simple, easily computable form of discord, diagonal
discord, and showed that the amount of diagonal discord
generated during heat transfer is directly proportional
to the heat flow between the two systems prepared at
two systems prepared at different transfer.   Elsewhere \cite{rd,2017arXiv170809076L},
we show that diagonal discord is a well-defined measure
of non-classicality: it is monotonically decreasing under
discord nongenerating transformations and is continuous
for generic states.   Here, we showed that diagonal
discord is equal to ordinary discord in the Gaussian,
high-temperature limit.   We performed a nanoscale
heat transfer experiment that -- combined with our
theoretical results -- allowed us to measure
macroscopic rates of creation of discord and diagonal
discord.     The connection between fundamental measures
of nonclassicality and thermodynamic processes suggests
that discord and diagonal discord may provide useful
tools for the analysis of quantum phenomena.
On a relevant note, the quantification of coherence and discord are rigorously studied under the resource theory framework in recent years \cite{baumgratz,levi,winteryang,rd,2016arXiv160902439S}.

\smallskip

\begin{acknowledgements}
This work was supported by DARPA,
ARO, AFOSR, the MIT Energy Initiative, and Jeffrey Epstein.
\end{acknowledgements}


\begin{thebibliography}{38}%
\makeatletter
\providecommand \@ifxundefined [1]{%
 \@ifx{#1\undefined}
}%
\providecommand \@ifnum [1]{%
 \ifnum #1\expandafter \@firstoftwo
 \else \expandafter \@secondoftwo
 \fi
}%
\providecommand \@ifx [1]{%
 \ifx #1\expandafter \@firstoftwo
 \else \expandafter \@secondoftwo
 \fi
}%
\providecommand \natexlab [1]{#1}%
\providecommand \enquote  [1]{``#1''}%
\providecommand \bibnamefont  [1]{#1}%
\providecommand \bibfnamefont [1]{#1}%
\providecommand \citenamefont [1]{#1}%
\providecommand \href@noop [0]{\@secondoftwo}%
\providecommand \href [0]{\begingroup \@sanitize@url \@href}%
\providecommand \@href[1]{\@@startlink{#1}\@@href}%
\providecommand \@@href[1]{\endgroup#1\@@endlink}%
\providecommand \@sanitize@url [0]{\catcode `\\12\catcode `\$12\catcode
  `\&12\catcode `\#12\catcode `\^12\catcode `\_12\catcode `\%12\relax}%
\providecommand \@@startlink[1]{}%
\providecommand \@@endlink[0]{}%
\providecommand \url  [0]{\begingroup\@sanitize@url \@url }%
\providecommand \@url [1]{\endgroup\@href {#1}{\urlprefix }}%
\providecommand \urlprefix  [0]{URL }%
\providecommand \Eprint [0]{\href }%
\providecommand \doibase [0]{http://dx.doi.org/}%
\providecommand \selectlanguage [0]{\@gobble}%
\providecommand \bibinfo  [0]{\@secondoftwo}%
\providecommand \bibfield  [0]{\@secondoftwo}%
\providecommand \translation [1]{[#1]}%
\providecommand \BibitemOpen [0]{}%
\providecommand \bibitemStop [0]{}%
\providecommand \bibitemNoStop [0]{.\EOS\space}%
\providecommand \EOS [0]{\spacefactor3000\relax}%
\providecommand \BibitemShut  [1]{\csname bibitem#1\endcsname}%
\let\auto@bib@innerbib\@empty
\bibitem [{\citenamefont {Nielsen}\ and\ \citenamefont {Chuang}(2000)}]{nc}%
  \BibitemOpen
  \bibfield  {author} {\bibinfo {author} {\bibfnamefont {M.~A.}\ \bibnamefont
  {Nielsen}}\ and\ \bibinfo {author} {\bibfnamefont {I.~L.}\ \bibnamefont
  {Chuang}},\ }\href@noop {} {\emph {\bibinfo {title} {Quantum Computation and
  Quantum Information}}}\ (\bibinfo  {publisher} {Cambridge University Press},\
  \bibinfo {year} {2000})\BibitemShut {NoStop}%
\bibitem [{\citenamefont {Ollivier}\ and\ \citenamefont {Zurek}(2001)}]{zurek}%
  \BibitemOpen
  \bibfield  {author} {\bibinfo {author} {\bibfnamefont {H.}~\bibnamefont
  {Ollivier}}\ and\ \bibinfo {author} {\bibfnamefont {W.~H.}\ \bibnamefont
  {Zurek}},\ }\href {\doibase 10.1103/PhysRevLett.88.017901} {\bibfield
  {journal} {\bibinfo  {journal} {Phys. Rev. Lett.}\ }\textbf {\bibinfo
  {volume} {88}},\ \bibinfo {pages} {017901} (\bibinfo {year}
  {2001})}\BibitemShut {NoStop}%
\bibitem [{\citenamefont {Zurek}(2003)}]{zurekmd}%
  \BibitemOpen
  \bibfield  {author} {\bibinfo {author} {\bibfnamefont {W.~H.}\ \bibnamefont
  {Zurek}},\ }\href {\doibase 10.1103/PhysRevA.67.012320} {\bibfield  {journal}
  {\bibinfo  {journal} {Phys. Rev. A}\ }\textbf {\bibinfo {volume} {67}},\
  \bibinfo {pages} {012320} (\bibinfo {year} {2003})}\BibitemShut {NoStop}%
\bibitem [{\citenamefont {Henderson}\ and\ \citenamefont {Vedral}(2001)}]{hv}%
  \BibitemOpen
  \bibfield  {author} {\bibinfo {author} {\bibfnamefont {L.}~\bibnamefont
  {Henderson}}\ and\ \bibinfo {author} {\bibfnamefont {V.}~\bibnamefont
  {Vedral}},\ }\href {http://stacks.iop.org/0305-4470/34/i=35/a=315} {\bibfield
   {journal} {\bibinfo  {journal} {Journal of Physics A: Mathematical and
  General}\ }\textbf {\bibinfo {volume} {34}},\ \bibinfo {pages} {6899}
  (\bibinfo {year} {2001})}\BibitemShut {NoStop}%
\bibitem [{\citenamefont {Oppenheim}\ \emph {et~al.}(2002)\citenamefont
  {Oppenheim}, \citenamefont {Horodecki}, \citenamefont {Horodecki},\ and\
  \citenamefont {Horodecki}}]{ohhh}%
  \BibitemOpen
  \bibfield  {author} {\bibinfo {author} {\bibfnamefont {J.}~\bibnamefont
  {Oppenheim}}, \bibinfo {author} {\bibfnamefont {M.}~\bibnamefont
  {Horodecki}}, \bibinfo {author} {\bibfnamefont {P.}~\bibnamefont
  {Horodecki}}, \ and\ \bibinfo {author} {\bibfnamefont {R.}~\bibnamefont
  {Horodecki}},\ }\href {\doibase 10.1103/PhysRevLett.89.180402} {\bibfield
  {journal} {\bibinfo  {journal} {Phys. Rev. Lett.}\ }\textbf {\bibinfo
  {volume} {89}},\ \bibinfo {pages} {180402} (\bibinfo {year}
  {2002})}\BibitemShut {NoStop}%
\bibitem [{\citenamefont {Piani}\ \emph {et~al.}(2008)\citenamefont {Piani},
  \citenamefont {Horodecki},\ and\ \citenamefont {Horodecki}}]{phh}%
  \BibitemOpen
  \bibfield  {author} {\bibinfo {author} {\bibfnamefont {M.}~\bibnamefont
  {Piani}}, \bibinfo {author} {\bibfnamefont {P.}~\bibnamefont {Horodecki}}, \
  and\ \bibinfo {author} {\bibfnamefont {R.}~\bibnamefont {Horodecki}},\ }\href
  {\doibase 10.1103/PhysRevLett.100.090502} {\bibfield  {journal} {\bibinfo
  {journal} {Phys. Rev. Lett.}\ }\textbf {\bibinfo {volume} {100}},\ \bibinfo
  {pages} {090502} (\bibinfo {year} {2008})}\BibitemShut {NoStop}%
\bibitem [{\citenamefont {Wu}\ \emph {et~al.}(2009)\citenamefont {Wu},
  \citenamefont {Poulsen},\ and\ \citenamefont {M\o{}lmer}}]{wpm}%
  \BibitemOpen
  \bibfield  {author} {\bibinfo {author} {\bibfnamefont {S.}~\bibnamefont
  {Wu}}, \bibinfo {author} {\bibfnamefont {U.~V.}\ \bibnamefont {Poulsen}}, \
  and\ \bibinfo {author} {\bibfnamefont {K.}~\bibnamefont {M\o{}lmer}},\ }\href
  {\doibase 10.1103/PhysRevA.80.032319} {\bibfield  {journal} {\bibinfo
  {journal} {Phys. Rev. A}\ }\textbf {\bibinfo {volume} {80}},\ \bibinfo
  {pages} {032319} (\bibinfo {year} {2009})}\BibitemShut {NoStop}%
\bibitem [{\citenamefont {Modi}\ \emph {et~al.}(2010)\citenamefont {Modi},
  \citenamefont {Paterek}, \citenamefont {Son}, \citenamefont {Vedral},\ and\
  \citenamefont {Williamson}}]{mpsvw}%
  \BibitemOpen
  \bibfield  {author} {\bibinfo {author} {\bibfnamefont {K.}~\bibnamefont
  {Modi}}, \bibinfo {author} {\bibfnamefont {T.}~\bibnamefont {Paterek}},
  \bibinfo {author} {\bibfnamefont {W.}~\bibnamefont {Son}}, \bibinfo {author}
  {\bibfnamefont {V.}~\bibnamefont {Vedral}}, \ and\ \bibinfo {author}
  {\bibfnamefont {M.}~\bibnamefont {Williamson}},\ }\href {\doibase
  10.1103/PhysRevLett.104.080501} {\bibfield  {journal} {\bibinfo  {journal}
  {Phys. Rev. Lett.}\ }\textbf {\bibinfo {volume} {104}},\ \bibinfo {pages}
  {080501} (\bibinfo {year} {2010})}\BibitemShut {NoStop}%
\bibitem [{\citenamefont {Datta}(2008)}]{datta}%
  \BibitemOpen
  \bibfield  {author} {\bibinfo {author} {\bibfnamefont {A.}~\bibnamefont
  {Datta}},\ }\emph {\bibinfo {title} {Studies on the Role of Entanglement in
  Mixed-State Quantum Computation}},\ \href@noop {} {Ph.D. thesis},\ \bibinfo
  {school} {University of New Mexico}, \bibinfo {address} {Albuquerque, New
  Mexico} (\bibinfo {year} {2008})\BibitemShut {NoStop}%
\bibitem [{\citenamefont {Datta}\ and\ \citenamefont {Gharibian}(2009)}]{dg}%
  \BibitemOpen
  \bibfield  {author} {\bibinfo {author} {\bibfnamefont {A.}~\bibnamefont
  {Datta}}\ and\ \bibinfo {author} {\bibfnamefont {S.}~\bibnamefont
  {Gharibian}},\ }\href {\doibase 10.1103/PhysRevA.79.042325} {\bibfield
  {journal} {\bibinfo  {journal} {Phys. Rev. A}\ }\textbf {\bibinfo {volume}
  {79}},\ \bibinfo {pages} {042325} (\bibinfo {year} {2009})}\BibitemShut
  {NoStop}%
\bibitem [{\citenamefont {Luo}(2008{\natexlab{a}})}]{luo}%
  \BibitemOpen
  \bibfield  {author} {\bibinfo {author} {\bibfnamefont {S.}~\bibnamefont
  {Luo}},\ }\href {\doibase 10.1103/PhysRevA.77.022301} {\bibfield  {journal}
  {\bibinfo  {journal} {Phys. Rev. A}\ }\textbf {\bibinfo {volume} {77}},\
  \bibinfo {pages} {022301} (\bibinfo {year} {2008}{\natexlab{a}})}\BibitemShut
  {NoStop}%
\bibitem [{\citenamefont {Hayden}\ \emph {et~al.}(2004)\citenamefont {Hayden},
  \citenamefont {Jozsa}, \citenamefont {Petz},\ and\ \citenamefont
  {Winter}}]{hjpw}%
  \BibitemOpen
  \bibfield  {author} {\bibinfo {author} {\bibfnamefont {P.}~\bibnamefont
  {Hayden}}, \bibinfo {author} {\bibfnamefont {R.}~\bibnamefont {Jozsa}},
  \bibinfo {author} {\bibfnamefont {D.}~\bibnamefont {Petz}}, \ and\ \bibinfo
  {author} {\bibfnamefont {A.}~\bibnamefont {Winter}},\ }\href {\doibase
  10.1007/s00220-004-1049-z} {\bibfield  {journal} {\bibinfo  {journal}
  {Communications in Mathematical Physics}\ }\textbf {\bibinfo {volume}
  {246}},\ \bibinfo {pages} {359} (\bibinfo {year} {2004})}\BibitemShut
  {NoStop}%
\bibitem [{\citenamefont {Bennett}\ \emph {et~al.}(1999)\citenamefont
  {Bennett}, \citenamefont {DiVincenzo}, \citenamefont {Fuchs}, \citenamefont
  {Mor}, \citenamefont {Rains}, \citenamefont {Shor}, \citenamefont {Smolin},\
  and\ \citenamefont {Wootters}}]{bennett}%
  \BibitemOpen
  \bibfield  {author} {\bibinfo {author} {\bibfnamefont {C.~H.}\ \bibnamefont
  {Bennett}}, \bibinfo {author} {\bibfnamefont {D.~P.}\ \bibnamefont
  {DiVincenzo}}, \bibinfo {author} {\bibfnamefont {C.~A.}\ \bibnamefont
  {Fuchs}}, \bibinfo {author} {\bibfnamefont {T.}~\bibnamefont {Mor}}, \bibinfo
  {author} {\bibfnamefont {E.}~\bibnamefont {Rains}}, \bibinfo {author}
  {\bibfnamefont {P.~W.}\ \bibnamefont {Shor}}, \bibinfo {author}
  {\bibfnamefont {J.~A.}\ \bibnamefont {Smolin}}, \ and\ \bibinfo {author}
  {\bibfnamefont {W.~K.}\ \bibnamefont {Wootters}},\ }\href {\doibase
  10.1103/PhysRevA.59.1070} {\bibfield  {journal} {\bibinfo  {journal} {Phys.
  Rev. A}\ }\textbf {\bibinfo {volume} {59}},\ \bibinfo {pages} {1070}
  (\bibinfo {year} {1999})}\BibitemShut {NoStop}%
\bibitem [{\citenamefont {{Lang}}\ \emph {et~al.}(2011)\citenamefont {{Lang}},
  \citenamefont {{Caves}},\ and\ \citenamefont {{Shaji}}}]{lcs}%
  \BibitemOpen
  \bibfield  {author} {\bibinfo {author} {\bibfnamefont {M.~D.}\ \bibnamefont
  {{Lang}}}, \bibinfo {author} {\bibfnamefont {C.~M.}\ \bibnamefont {{Caves}}},
  \ and\ \bibinfo {author} {\bibfnamefont {A.}~\bibnamefont {{Shaji}}},\
  }\href@noop {} {\bibfield  {journal} {\bibinfo  {journal} {ArXiv e-prints}\ }
  (\bibinfo {year} {2011})},\ \Eprint {http://arxiv.org/abs/1105.4920}
  {arXiv:1105.4920 [quant-ph]} \BibitemShut {NoStop}%
\bibitem [{\citenamefont {Wu}\ and\ \citenamefont {Segal}(2011)}]{ws}%
  \BibitemOpen
  \bibfield  {author} {\bibinfo {author} {\bibfnamefont {L.-A.}\ \bibnamefont
  {Wu}}\ and\ \bibinfo {author} {\bibfnamefont {D.}~\bibnamefont {Segal}},\
  }\href {\doibase 10.1103/PhysRevA.84.012319} {\bibfield  {journal} {\bibinfo
  {journal} {Phys. Rev. A}\ }\textbf {\bibinfo {volume} {84}},\ \bibinfo
  {pages} {012319} (\bibinfo {year} {2011})}\BibitemShut {NoStop}%
\bibitem [{\citenamefont {Giovannetti}\ \emph {et~al.}(2004)\citenamefont
  {Giovannetti}, \citenamefont {Lloyd},\ and\ \citenamefont {Maccone}}]{glm}%
  \BibitemOpen
  \bibfield  {author} {\bibinfo {author} {\bibfnamefont {V.}~\bibnamefont
  {Giovannetti}}, \bibinfo {author} {\bibfnamefont {S.}~\bibnamefont {Lloyd}},
  \ and\ \bibinfo {author} {\bibfnamefont {L.}~\bibnamefont {Maccone}},\ }\href
  {\doibase 10.1126/science.1104149} {\bibfield  {journal} {\bibinfo  {journal}
  {Science}\ }\textbf {\bibinfo {volume} {306}},\ \bibinfo {pages} {1330}
  (\bibinfo {year} {2004})}\BibitemShut {NoStop}%
\bibitem [{\citenamefont {Rebentrost}\ \emph {et~al.}(2009)\citenamefont
  {Rebentrost}, \citenamefont {Mohseni}, \citenamefont {Kassal}, \citenamefont
  {Lloyd},\ and\ \citenamefont {Aspuru-Guzik}}]{reben}%
  \BibitemOpen
  \bibfield  {author} {\bibinfo {author} {\bibfnamefont {P.}~\bibnamefont
  {Rebentrost}}, \bibinfo {author} {\bibfnamefont {M.}~\bibnamefont {Mohseni}},
  \bibinfo {author} {\bibfnamefont {I.}~\bibnamefont {Kassal}}, \bibinfo
  {author} {\bibfnamefont {S.}~\bibnamefont {Lloyd}}, \ and\ \bibinfo {author}
  {\bibfnamefont {A.}~\bibnamefont {Aspuru-Guzik}},\ }\href
  {http://stacks.iop.org/1367-2630/11/i=3/a=033003} {\bibfield  {journal}
  {\bibinfo  {journal} {New Journal of Physics}\ }\textbf {\bibinfo {volume}
  {11}},\ \bibinfo {pages} {033003} (\bibinfo {year} {2009})}\BibitemShut
  {NoStop}%
\bibitem [{\citenamefont {Plenio}\ and\ \citenamefont {Huelga}(2008)}]{ph}%
  \BibitemOpen
  \bibfield  {author} {\bibinfo {author} {\bibfnamefont {M.~B.}\ \bibnamefont
  {Plenio}}\ and\ \bibinfo {author} {\bibfnamefont {S.~F.}\ \bibnamefont
  {Huelga}},\ }\href {http://stacks.iop.org/1367-2630/10/i=11/a=113019}
  {\bibfield  {journal} {\bibinfo  {journal} {New Journal of Physics}\ }\textbf
  {\bibinfo {volume} {10}},\ \bibinfo {pages} {113019} (\bibinfo {year}
  {2008})}\BibitemShut {NoStop}%
\bibitem [{\citenamefont {{Lloyd}}\ \emph {et~al.}(2011)\citenamefont
  {{Lloyd}}, \citenamefont {{Mohseni}}, \citenamefont {{Shabani}},\ and\
  \citenamefont {{Rabitz}}}]{goldi}%
  \BibitemOpen
  \bibfield  {author} {\bibinfo {author} {\bibfnamefont {S.}~\bibnamefont
  {{Lloyd}}}, \bibinfo {author} {\bibfnamefont {M.}~\bibnamefont {{Mohseni}}},
  \bibinfo {author} {\bibfnamefont {A.}~\bibnamefont {{Shabani}}}, \ and\
  \bibinfo {author} {\bibfnamefont {H.}~\bibnamefont {{Rabitz}}},\ }\href@noop
  {} {\bibfield  {journal} {\bibinfo  {journal} {ArXiv e-prints}\ } (\bibinfo
  {year} {2011})},\ \Eprint {http://arxiv.org/abs/1111.4982} {arXiv:1111.4982
  [quant-ph]} \BibitemShut {NoStop}%
\bibitem [{\citenamefont {Pirandola}(2014)}]{Pirandola2014}%
  \BibitemOpen
  \bibfield  {author} {\bibinfo {author} {\bibfnamefont {S.}~\bibnamefont
  {Pirandola}},\ }\href {http://dx.doi.org/10.1038/srep06956
  http://10.0.4.14/srep06956
  https://www.nature.com/articles/srep06956{\#}supplementary-information}
  {\bibfield  {journal} {\bibinfo  {journal} {Sci. Rep.}\ }\textbf {\bibinfo
  {volume} {4}},\ \bibinfo {pages} {6956} (\bibinfo {year} {2014})}\BibitemShut
  {NoStop}%
\bibitem [{\citenamefont {Pirandola}\ \emph {et~al.}(2017)\citenamefont
  {Pirandola}, \citenamefont {Laurenza}, \citenamefont {Ottaviani},\ and\
  \citenamefont {Banchi}}]{Pirandola2017}%
  \BibitemOpen
  \bibfield  {author} {\bibinfo {author} {\bibfnamefont {S.}~\bibnamefont
  {Pirandola}}, \bibinfo {author} {\bibfnamefont {R.}~\bibnamefont {Laurenza}},
  \bibinfo {author} {\bibfnamefont {C.}~\bibnamefont {Ottaviani}}, \ and\
  \bibinfo {author} {\bibfnamefont {L.}~\bibnamefont {Banchi}},\ }\href
  {http://dx.doi.org/10.1038/ncomms15043 http://10.0.4.14/ncomms15043
  https://www.nature.com/articles/ncomms15043{\#}supplementary-information}
  {\bibfield  {journal} {\bibinfo  {journal} {Nat. Commun.}\ }\textbf {\bibinfo
  {volume} {8}},\ \bibinfo {pages} {15043} (\bibinfo {year}
  {2017})}\BibitemShut {NoStop}%
\bibitem [{\citenamefont {Modi}\ \emph {et~al.}(2012)\citenamefont {Modi},
  \citenamefont {Brodutch}, \citenamefont {Cable}, \citenamefont {Paterek},\
  and\ \citenamefont {Vedral}}]{modi}%
  \BibitemOpen
  \bibfield  {author} {\bibinfo {author} {\bibfnamefont {K.}~\bibnamefont
  {Modi}}, \bibinfo {author} {\bibfnamefont {A.}~\bibnamefont {Brodutch}},
  \bibinfo {author} {\bibfnamefont {H.}~\bibnamefont {Cable}}, \bibinfo
  {author} {\bibfnamefont {T.}~\bibnamefont {Paterek}}, \ and\ \bibinfo
  {author} {\bibfnamefont {V.}~\bibnamefont {Vedral}},\ }\href {\doibase
  10.1103/RevModPhys.84.1655} {\bibfield  {journal} {\bibinfo  {journal} {Rev.
  Mod. Phys.}\ }\textbf {\bibinfo {volume} {84}},\ \bibinfo {pages} {1655}
  (\bibinfo {year} {2012})}\BibitemShut {NoStop}%
\bibitem [{\citenamefont {Huang}(2014)}]{npc}%
  \BibitemOpen
  \bibfield  {author} {\bibinfo {author} {\bibfnamefont {Y.}~\bibnamefont
  {Huang}},\ }\href {http://stacks.iop.org/1367-2630/16/i=3/a=033027}
  {\bibfield  {journal} {\bibinfo  {journal} {New Journal of Physics}\ }\textbf
  {\bibinfo {volume} {16}},\ \bibinfo {pages} {033027} (\bibinfo {year}
  {2014})}\BibitemShut {NoStop}%
\bibitem [{\citenamefont {Luo}(2008{\natexlab{b}})}]{mid}%
  \BibitemOpen
  \bibfield  {author} {\bibinfo {author} {\bibfnamefont {S.}~\bibnamefont
  {Luo}},\ }\href {\doibase 10.1103/PhysRevA.77.022301} {\bibfield  {journal}
  {\bibinfo  {journal} {Phys. Rev. A}\ }\textbf {\bibinfo {volume} {77}},\
  \bibinfo {pages} {022301} (\bibinfo {year} {2008}{\natexlab{b}})}\BibitemShut
  {NoStop}%
\bibitem [{\citenamefont {{Eastin}}(2010)}]{eastin}%
  \BibitemOpen
  \bibfield  {author} {\bibinfo {author} {\bibfnamefont {B.}~\bibnamefont
  {{Eastin}}},\ }\href@noop {} {\bibfield  {journal} {\bibinfo  {journal}
  {ArXiv e-prints}\ } (\bibinfo {year} {2010})},\ \Eprint
  {http://arxiv.org/abs/1006.4402} {arXiv:1006.4402 [quant-ph]} \BibitemShut
  {NoStop}%
\bibitem [{\citenamefont {Cable}\ and\ \citenamefont {Browne}(2015)}]{cable}%
  \BibitemOpen
  \bibfield  {author} {\bibinfo {author} {\bibfnamefont {H.}~\bibnamefont
  {Cable}}\ and\ \bibinfo {author} {\bibfnamefont {D.~E.}\ \bibnamefont
  {Browne}},\ }\href {http://stacks.iop.org/1367-2630/17/i=11/a=113049}
  {\bibfield  {journal} {\bibinfo  {journal} {New Journal of Physics}\ }\textbf
  {\bibinfo {volume} {17}},\ \bibinfo {pages} {113049} (\bibinfo {year}
  {2015})}\BibitemShut {NoStop}%
\bibitem [{\citenamefont {Brodutch}\ and\ \citenamefont
  {Modi}(2012)}]{criteria}%
  \BibitemOpen
  \bibfield  {author} {\bibinfo {author} {\bibfnamefont {A.}~\bibnamefont
  {Brodutch}}\ and\ \bibinfo {author} {\bibfnamefont {K.}~\bibnamefont
  {Modi}},\ }\href {http://dl.acm.org/citation.cfm?id=2481580.2481581}
  {\bibfield  {journal} {\bibinfo  {journal} {Quantum Info. Comput.}\ }\textbf
  {\bibinfo {volume} {12}},\ \bibinfo {pages} {721} (\bibinfo {year}
  {2012})}\BibitemShut {NoStop}%
\bibitem [{\citenamefont {{Liu}}\ \emph {et~al.}(2017)\citenamefont {{Liu}},
  \citenamefont {{Takagi}},\ and\ \citenamefont
  {{Lloyd}}}]{2017arXiv170809076L}%
  \BibitemOpen
  \bibfield  {author} {\bibinfo {author} {\bibfnamefont {Z.-W.}\ \bibnamefont
  {{Liu}}}, \bibinfo {author} {\bibfnamefont {R.}~\bibnamefont {{Takagi}}}, \
  and\ \bibinfo {author} {\bibfnamefont {S.}~\bibnamefont {{Lloyd}}},\
  }\href@noop {} {\bibfield  {journal} {\bibinfo  {journal} {ArXiv e-prints}\ }
  (\bibinfo {year} {2017})},\ \Eprint {http://arxiv.org/abs/1708.09076}
  {arXiv:1708.09076 [quant-ph]} \BibitemShut {NoStop}%
\bibitem [{\citenamefont {Liu}\ \emph {et~al.}(2017)\citenamefont {Liu},
  \citenamefont {Hu},\ and\ \citenamefont {Lloyd}}]{rd}%
  \BibitemOpen
  \bibfield  {author} {\bibinfo {author} {\bibfnamefont {Z.-W.}\ \bibnamefont
  {Liu}}, \bibinfo {author} {\bibfnamefont {X.}~\bibnamefont {Hu}}, \ and\
  \bibinfo {author} {\bibfnamefont {S.}~\bibnamefont {Lloyd}},\ }\href
  {\doibase 10.1103/PhysRevLett.118.060502} {\bibfield  {journal} {\bibinfo
  {journal} {Phys. Rev. Lett.}\ }\textbf {\bibinfo {volume} {118}},\ \bibinfo
  {pages} {060502} (\bibinfo {year} {2017})}\BibitemShut {NoStop}%
\bibitem [{\citenamefont {Pirandola}\ \emph {et~al.}(2014)\citenamefont
  {Pirandola}, \citenamefont {Spedalieri}, \citenamefont {Braunstein},
  \citenamefont {Cerf},\ and\ \citenamefont {Lloyd}}]{OptimalityDiscord}%
  \BibitemOpen
  \bibfield  {author} {\bibinfo {author} {\bibfnamefont {S.}~\bibnamefont
  {Pirandola}}, \bibinfo {author} {\bibfnamefont {G.}~\bibnamefont
  {Spedalieri}}, \bibinfo {author} {\bibfnamefont {S.~L.}\ \bibnamefont
  {Braunstein}}, \bibinfo {author} {\bibfnamefont {N.~J.}\ \bibnamefont
  {Cerf}}, \ and\ \bibinfo {author} {\bibfnamefont {S.}~\bibnamefont {Lloyd}},\
  }\href {\doibase 10.1103/PhysRevLett.113.140405} {\bibfield  {journal}
  {\bibinfo  {journal} {Phys. Rev. Lett.}\ }\textbf {\bibinfo {volume} {113}},\
  \bibinfo {pages} {140405} (\bibinfo {year} {2014})}\BibitemShut {NoStop}%
\bibitem [{\citenamefont {Brodutch}\ and\ \citenamefont
  {Terno}(2010)}]{dddemon}%
  \BibitemOpen
  \bibfield  {author} {\bibinfo {author} {\bibfnamefont {A.}~\bibnamefont
  {Brodutch}}\ and\ \bibinfo {author} {\bibfnamefont {D.~R.}\ \bibnamefont
  {Terno}},\ }\href {\doibase 10.1103/PhysRevA.81.062103} {\bibfield  {journal}
  {\bibinfo  {journal} {Phys. Rev. A}\ }\textbf {\bibinfo {volume} {81}},\
  \bibinfo {pages} {062103} (\bibinfo {year} {2010})}\BibitemShut {NoStop}%
\bibitem [{\citenamefont {Volokitin}\ and\ \citenamefont {Persson}(2007)}]{vp}%
  \BibitemOpen
  \bibfield  {author} {\bibinfo {author} {\bibfnamefont {A.~I.}\ \bibnamefont
  {Volokitin}}\ and\ \bibinfo {author} {\bibfnamefont {B.~N.~J.}\ \bibnamefont
  {Persson}},\ }\href {\doibase 10.1103/RevModPhys.79.1291} {\bibfield
  {journal} {\bibinfo  {journal} {Rev. Mod. Phys.}\ }\textbf {\bibinfo {volume}
  {79}},\ \bibinfo {pages} {1291} (\bibinfo {year} {2007})}\BibitemShut
  {NoStop}%
\bibitem [{\citenamefont {Cahill}\ \emph {et~al.}(2014)\citenamefont {Cahill},
  \citenamefont {Braun}, \citenamefont {Chen}, \citenamefont {Clarke},
  \citenamefont {Fan}, \citenamefont {Goodson}, \citenamefont {Keblinski},
  \citenamefont {King}, \citenamefont {Mahan}, \citenamefont {Majumdar},
  \citenamefont {Maris}, \citenamefont {Phillpot}, \citenamefont {Pop},\ and\
  \citenamefont {Shi}}]{trev}%
  \BibitemOpen
  \bibfield  {author} {\bibinfo {author} {\bibfnamefont {D.~G.}\ \bibnamefont
  {Cahill}}, \bibinfo {author} {\bibfnamefont {P.~V.}\ \bibnamefont {Braun}},
  \bibinfo {author} {\bibfnamefont {G.}~\bibnamefont {Chen}}, \bibinfo {author}
  {\bibfnamefont {D.~R.}\ \bibnamefont {Clarke}}, \bibinfo {author}
  {\bibfnamefont {S.}~\bibnamefont {Fan}}, \bibinfo {author} {\bibfnamefont
  {K.~E.}\ \bibnamefont {Goodson}}, \bibinfo {author} {\bibfnamefont
  {P.}~\bibnamefont {Keblinski}}, \bibinfo {author} {\bibfnamefont {W.~P.}\
  \bibnamefont {King}}, \bibinfo {author} {\bibfnamefont {G.~D.}\ \bibnamefont
  {Mahan}}, \bibinfo {author} {\bibfnamefont {A.}~\bibnamefont {Majumdar}},
  \bibinfo {author} {\bibfnamefont {H.~J.}\ \bibnamefont {Maris}}, \bibinfo
  {author} {\bibfnamefont {S.~R.}\ \bibnamefont {Phillpot}}, \bibinfo {author}
  {\bibfnamefont {E.}~\bibnamefont {Pop}}, \ and\ \bibinfo {author}
  {\bibfnamefont {L.}~\bibnamefont {Shi}},\ }\href {\doibase
  http://dx.doi.org/10.1063/1.4832615} {\bibfield  {journal} {\bibinfo
  {journal} {Applied Physics Reviews}\ }\textbf {\bibinfo {volume} {1}},\
  \bibinfo {eid} {011305} (\bibinfo {year} {2014})}\BibitemShut {NoStop}%
\bibitem [{\citenamefont {Baumgratz}\ \emph {et~al.}(2014)\citenamefont
  {Baumgratz}, \citenamefont {Cramer},\ and\ \citenamefont
  {Plenio}}]{baumgratz}%
  \BibitemOpen
  \bibfield  {author} {\bibinfo {author} {\bibfnamefont {T.}~\bibnamefont
  {Baumgratz}}, \bibinfo {author} {\bibfnamefont {M.}~\bibnamefont {Cramer}}, \
  and\ \bibinfo {author} {\bibfnamefont {M.~B.}\ \bibnamefont {Plenio}},\
  }\href {\doibase 10.1103/PhysRevLett.113.140401} {\bibfield  {journal}
  {\bibinfo  {journal} {Phys. Rev. Lett.}\ }\textbf {\bibinfo {volume} {113}},\
  \bibinfo {pages} {140401} (\bibinfo {year} {2014})}\BibitemShut {NoStop}%
\bibitem [{\citenamefont {Levi}\ and\ \citenamefont {Mintert}(2014)}]{levi}%
  \BibitemOpen
  \bibfield  {author} {\bibinfo {author} {\bibfnamefont {F.}~\bibnamefont
  {Levi}}\ and\ \bibinfo {author} {\bibfnamefont {F.}~\bibnamefont {Mintert}},\
  }\href {http://stacks.iop.org/1367-2630/16/i=3/a=033007} {\bibfield
  {journal} {\bibinfo  {journal} {New Journal of Physics}\ }\textbf {\bibinfo
  {volume} {16}},\ \bibinfo {pages} {033007} (\bibinfo {year}
  {2014})}\BibitemShut {NoStop}%
\bibitem [{\citenamefont {{Winter}}\ and\ \citenamefont
  {{Yang}}(2015)}]{winteryang}%
  \BibitemOpen
  \bibfield  {author} {\bibinfo {author} {\bibfnamefont {A.}~\bibnamefont
  {{Winter}}}\ and\ \bibinfo {author} {\bibfnamefont {D.}~\bibnamefont
  {{Yang}}},\ }\href@noop {} {\bibfield  {journal} {\bibinfo  {journal} {ArXiv
  e-prints}\ } (\bibinfo {year} {2015})},\ \Eprint
  {http://arxiv.org/abs/1506.07975} {arXiv:1506.07975 [quant-ph]} \BibitemShut
  {NoStop}%
\bibitem [{\citenamefont {{Streltsov}}\ \emph {et~al.}(2016)\citenamefont
  {{Streltsov}}, \citenamefont {{Adesso}},\ and\ \citenamefont
  {{Plenio}}}]{2016arXiv160902439S}%
  \BibitemOpen
  \bibfield  {author} {\bibinfo {author} {\bibfnamefont {A.}~\bibnamefont
  {{Streltsov}}}, \bibinfo {author} {\bibfnamefont {G.}~\bibnamefont
  {{Adesso}}}, \ and\ \bibinfo {author} {\bibfnamefont {M.~B.}\ \bibnamefont
  {{Plenio}}},\ }\href@noop {} {\bibfield  {journal} {\bibinfo  {journal}
  {ArXiv e-prints}\ } (\bibinfo {year} {2016})},\ \Eprint
  {http://arxiv.org/abs/1609.02439} {arXiv:1609.02439 [quant-ph]} \BibitemShut
  {NoStop}%
\bibitem [{\citenamefont {Adesso}\ and\ \citenamefont
  {Datta}(2010)}]{adessodatta}%
  \BibitemOpen
  \bibfield  {author} {\bibinfo {author} {\bibfnamefont {G.}~\bibnamefont
  {Adesso}}\ and\ \bibinfo {author} {\bibfnamefont {A.}~\bibnamefont {Datta}},\
  }\href {\doibase 10.1103/PhysRevLett.105.030501} {\bibfield  {journal}
  {\bibinfo  {journal} {Phys. Rev. Lett.}\ }\textbf {\bibinfo {volume} {105}},\
  \bibinfo {pages} {030501} (\bibinfo {year} {2010})}\BibitemShut {NoStop}%
\end{thebibliography}

%


\appendix
\begin{center}
\textbf{Supplementary Material}
\end{center}

\section{Gaussian discord for the heat transfer model}\label{gau}

Let us consider a thermal state at temperature $T$, i.e.,
\begin{equation}
\rho=\frac{e^{-\beta\hat{a}^{\dagger}\hat{a}}}{Z}=\frac{e^{-\beta(\hat{q}%
^{2}+\hat{p}^{2})/2}}{Z},
\end{equation}
where $\beta=(kT)^{-1}$. The mean photon number is
\begin{equation}
\bar{n}=\frac{1}{e^{\beta}-1},~~~\beta=\ln\left(  \frac{1+\bar{n}}{\bar{n}%
}\right)  ~.
\end{equation}
Its covariance matrix (CM) is given by $\mathbf{V}=\mu\mathbf{I}$, where
$\mathbf{I}$ is the identity and $\mu=\bar{n}+1/2$. Its entropy is
$S(\rho)=h(\mu)$, with%
\begin{equation}
h(\mu)=\left(  \mu+\frac{1}{2}\right)  \ln\left(  \mu+\frac{1}{2}\right)
-\left(  \mu-\frac{1}{2}\right)  \ln\left(  \mu-\frac{1}{2}\right)
~~\text{nats}.
\end{equation}

Consider two modes $A$ and $B$ in thermal states at temperatures $T_{A}>T_{B}%
$. Let us evolve them through a beam splitter (the most obvious passive
Gaussian interaction) with transmissivity $\eta$. We realise weak coupling for
$\eta\simeq1$. The input state $\rho_{AB}=\rho_{A}\otimes\rho_{B}$ with CM
$a\mathbf{I}\oplus b\mathbf{I}$ is transformed into the output
correlated-thermal state $\rho_{A^{\prime}B^{\prime}}$ with CM%
\begin{equation}
\mathbf{V}_{A^{\prime}B^{\prime}}=\left(
\begin{array}
[c]{cc}%
a^{\prime}\mathbf{I} & c^{\prime}\mathbf{I}\\
c^{\prime}\mathbf{I} & b^{\prime}\mathbf{I}%
\end{array}
\right)  ,\label{cm}%
\end{equation}
with%
\begin{align}
a^{\prime} &  :=\eta a+(1-\eta)b\\
b^{\prime} &  :=\eta b+(1-\eta)a\\
c^{\prime} &  :=\sqrt{\eta(1-\eta)}(b-a).
\end{align}
Note that, for the mean photon numbers, we may write
\begin{align}
\bar{n}_{A^{\prime}} &  :=\eta\bar{n}_{A}+(1-\eta)\bar{n}_{B}\\
\bar{n}_{B^{\prime}} &  :=\eta\bar{n}_{B}+(1-\eta)\bar{n}_{A}~.
\end{align}
Change in energy is proportional to%
\begin{align}
\Delta\bar{n}_{B} &  :=\bar{n}_{B^{\prime}}-\bar{n}_{B}=(1-\eta)(\bar{n}%
_{A}-\bar{n}_{B})>0~,\\
\Delta\bar{n}_{A} &  :=-\Delta\bar{n}_{B}~.
\end{align}
The symplectic eigenvalues of the CM~(\ref{cm}) are simply $a$ and $b$.

Note that for Gaussian states with CM as in Eq.~(\ref{cm}), we
proved~\cite{OptimalityDiscord} that quantum discord $D(B\rightarrow A)$ is
equal to the Gaussian discord $D_{G}(B\rightarrow A)$, where the measurement
on $B$ is restricted to a Gaussian POVM. Then, we may use the formula given in
Ref.~\cite{adessodatta}\ for the Gaussian discord. We find%
\begin{equation}
D(B\rightarrow A)=h(b^{\prime})-h(a)-h(b)+h\left(  \sqrt{e_{\min}}\right)  ,
\end{equation}
where $e_{\min}$ can be adapted from Eq.~(4) of Ref.~\cite{adessodatta}. 
The optimal measurement
achieving $D(B\rightarrow A)$ is heterodyne detection in this case.

At the first order in $\eta\simeq1$, we get%
\begin{align}
D(B &  \rightarrow A)=\Delta\bar{n}_{B}\left[  \ln\left(  \frac{1+\bar{n}_{B}%
}{\bar{n}_{B}}\right)  -\frac{1+\bar{n}_{A}}{1+\bar{n}_{B}}\ln\left(
\frac{1+\bar{n}_{A}}{\bar{n}_{A}}\right)  \right]  \nonumber\\
&  =\Delta\bar{n}_{B}\left[  \frac{1}{kT_{B}}-\frac{1+\bar{n}_{A}}{1+\bar
{n}_{B}}\frac{1}{kT_{A}}\right]  \\
&  \leq\Delta\bar{n}_{B}\left(  \frac{1}{kT_{B}}-\frac{1}{kT_{A}}\right)
=D_{diag}(B\rightarrow A)~.
\end{align}

\section{A simple spin model}
A simple example that illustrates our result is given as follows.
Consider two two-level systems such as spins with Hamiltonian
$H = (\hbar/2)\big( - \omega \sigma_z^A - \omega \sigma_z^B
+ \gamma(\sigma_+^A\otimes \sigma_-^B + \sigma_-^A\otimes\sigma_+^B) \big)$.
This is a simple swapping Hamiltonian that induces the continuous
time tunneling evolution
\begin{equation}\label{exp}
\begin{split}
&|0\rangle_A|0\rangle_B \rightarrow e^{i\omega t}|0\rangle_A|0\rangle_B\\
&|1\rangle_A|1\rangle_B
\rightarrow e^{-i\omega t} |1\rangle_A|1\rangle_B\\
&|0\rangle_A|1\rangle_B
\rightarrow \cos \gamma t/2 |0\rangle_A|1\rangle_B
-i \sin\gamma t/2 |1\rangle_A |0\rangle_B\\
&|1\rangle_A|0\rangle_B
\rightarrow \cos \gamma t/2 |1\rangle_A|0\rangle_B
-i \sin\gamma t/2 |1\rangle_A |0\rangle_B
\end{split}
\end{equation}

The spins are prepared in thermal states at different temperatures,
so that $\rho_{AB}(0) =
Z_A^{-1} e^{\omega \sigma_z^A/kT_A} \otimes
Z_B^{-1} e^{\omega \sigma_z^B/kT_B}$.
The energy transfer from $A$ to $B$ is due to the continuous
swapping within the $01$, $10$ subspace.
It is straightforward to verify that the initial energy transfer over
time $\Delta t$ obeys Eq.~(\ref{flow}).

The continuous tunneling dynamics implies that the amount of
energy transfer is proportional to $\sin^2 \gamma t/2$, which
reaches its maximum rate at $\gamma t = \pi/2$, halfway through
the tunneling process.  Because of the sinusoidal nature of
the tunneling dynamics, the energetic discord also rises to
its maximum value at this point.  The energetic discord
then decreases, reaching zero
when $\gamma t = \pi$, the point at which the states
of $A$ and $B$ have been swapped, so that $B$ is in a thermal
state at temperature $T_A$ and $A$ is in a thermal state
at temperature $T_B$.

\end{document}